\documentclass[12pt]{iopart}

\usepackage{graphicx}

\begin{document}

\title[]{Exact annihilation energy and proper decay time solution of a para-positronium system}

\author{Abdullah Guvendi and Yusuf Sucu}

\address{Department of Physics, Faculty of Science, Akdeniz University, TR-07058 Antalya, Turkey}
\ead{abdullahguvendi@gmail.com and ysucu@akdeniz.edu.tr}

\vspace{10pt}

\begin{abstract}
Para positronium composed by an electron-antielectron pair is an unstable system decaying into two high energetic gamma photons via self annihilation process, due to the conservation of the charge conjugation parity in electromagnetically interacting systems. Therefore, the spectrum covering all fundamental properties of the para-positronium system includes an imaginary part corresponding to the proper decay time besides the real parts corresponding to the total annihilation energy and binding energy, simultaneously. The para-positronium can be regarded as relativistic two body system in which there exist a Coulomb interaction force between the oppositely charged particles. Because of the annihilation condition, ($l=0$), and total spin of the system, ($S=0$), the problem is solved in $1+1$ dimensional spacetime background by using fully covariant relativistic two body equation, without any approximation. Adopting the obtained spectra to an electron-antielectron pair we find total annihilation energy, binding energy and proper decay time of the para-positronium system. Since the obtained spectra shows the fascinating properties of the system, our findings can shed light to medical monitoring processes, positron annihilation spectroscopy in any system and gamma-ray laser studies.
\end{abstract}

%
%
%
%

\section{Introduction}

One of the fundamental problems of the quantum electrodynamics is Positronium (Ps) atom \cite{c} which is bound state of an electron and an anti-electron. When the system is in the ground state ($n=1$), the Ps atom can be formed by spin symmetric, $1^{3}S_{1}$, or spin antisymmetric, $1^{1}S_{0}$, quantum states. The both quantum state of the Ps atom known as ortho-positronium (o-Ps, $1^{3}S_{1}$) and para-positronium (p-Ps, $1^{1}S_{0}$) decay via self-annihilation process by emitting high energetic gamma photons, eventually. Because of the conservation of the overall charge conjugation parity, ($-(-1)^{l+s}$), in electromagnetically interacting systems, the p-Ps system decays into two gamma photon propagating into opposite direction since the charge conjugation parity of a photon is odd ($-1$). Also, due to the binding energy (negative) between the oppositely charged particles, the total energy transmitting by the two annihilation photon (annihilation energy) must be lower than the total rest mass energy of the system ($2m_{e}c^{2}$).

\qquad Due to the Coulomb interaction between the two oppositely charged elementary particles, the both quantum state of Ps system are collapsed and annihilation condition, which means the annihilation event occurs only when the wave functions of the particles are overlapped \cite{d,e} at origin of the center of mass frame, is satisfied. However, the observed lifetimes of the both state differ from each other because of the conservation of the total angular momentum in their collapsing processes, spin-spin interaction force between the particles, phase space and additional alpha supression factor \cite{g,b1}. The observed total lifetimes ($\tau_{lab}$) of p-Ps and o-Ps systems are approximately $125\times10^{-12}$ and $142\times10^{-9}$ seconds \cite{b1,Ramadhan}, respectively, without any external effect \cite{e,Sa}, such as substrate effect or screening energy causing from the electronic environment of the annihilation position. Since the both system are formed by oppositely charged two fermion, the external effects make their collapsing processes delay and their binding energies changes and thus the effects increase their total annihilation times. It is useful to emphasize that, after positron emitted by a radioactive source, it losses their kinetic energy in some physical processes and binds the nearest an electron in the sample. Therefore, the annihilation event can be investigated in event-by-event basis. It can be seen that the average annihilation lifetime spectroscopy of the positrons in living biological systems shows differences in picosecond order between normal and cancerous cells, because of the physical and chemical processes are carrying out inside the medium of cells or tissues \cite{axpe,cancerous} and some morphological alterations in the free voids between the atoms of the sample matter in nanometer scale. Therefore, by measuring the mean annihilation lifetime and the total annihilation energy transmitting by the high energetic photons originating from the both annihilation process or by measuring the Doppler broadenings \cite{Asoka} of the resulting rays some physical informations about the affected cell membranes and tissues in living biological systems or the interested materials can be dedected since the annihilation photons hold information about the electronic environment around the annihilation point and morphology of the interested sample, via positron emission tomography (PET) and multi-purpose J-PET whole body scanners \cite{M,N,O} based on multi-photon registration. Therefore, the separation of the proper decay time values of the quantum states of Ps from the observed mean life time values is necessary to clarify the findings as well as the both total annihilation energy and binding energy changing with respect to the electronic properties of the medium. Therefore, firstly there must exist a spectrum including all fundamental properties of the p-Ps system as we mentioned before.

Hence, the fundamental aim of this research is to perform an exact solution  including simultaneously the total annihilation energy, binding energy and proper decay time of a p-Ps system, by excluding all external effects on the system. To achieve this, for the quantum state of p-Ps ($s=0,l=0$), the fully covariant relativistic two-body equation \cite{j,abdullah} is written in the $(1+1)$ dimensional background, and then relative and center of mass variables are separated by excepting the center of mass is rest. By taking $m_{1}=m_{2}$, the obtained second order wave equation is solved for the p-Ps system where the both particle hold together by a Coulomb interaction. Then, by modifying the obtained spectra to an electron-antielectron system, as we expected, the exact total annihilation energy, binding energy and proper decay time of the p-Ps is found, simultaneously. Furthermore, it can seen that our results can be adopted to any medium.

\section{Covariant two-body equation in (1+1) dimensions}

In $(1+1)$ dimensional space-times, the relativistic two-body equation can be written for the p-Ps system as \cite{abdullah,j};
\begin{eqnarray}
[(\sigma _{\mu }^{(1)}i\hbar (\partial_{\mu }^{(1) }+\frac{ie_{1}}{\hbar c}A_{\mu }^{(2) })
-m_{1}cI) \otimes \sigma _{0}^{(2) }\nonumber \\+\sigma _{0}^{(1)}\otimes ( \sigma _{\mu }^{(2)}i\hbar (
\partial _{\mu }^{(2)}+\frac{ie_{2}}{\hbar c}A_{\mu }^{(1)})-m_{2}cI)] \Psi( \mathbf{x}_{1},\mathbf{x}_{2}) =0,\nonumber\\
\partial _{\mu }=\left( \frac{\partial _{t}}{c},\partial _{x}\right),\quad \left(\mu=0,1.\right), \label{Eq1}
\end{eqnarray}
where, bi-spinor function is
\begin{eqnarray}
\Psi \left(\mathbf{x}_{1},\mathbf{x}_{2}\right) =\psi \left( \mathbf{x}_{1}\right) \otimes \psi \left( \mathbf{x}_{2}\right),\nonumber \\
\psi \left( \mathbf{x}_{1}\right)=\ \left(\begin{array}{c}\kappa\left( \mathbf{x}_{1}\right)\\ \chi\left( \mathbf{x}_{1}\right)\\ \end{array}\right), \psi \left( \mathbf{x}_{2}\right)=\ \left(\begin{array}{c}\kappa\left( \mathbf{x}_{2}\right)\\ \chi\left( \mathbf{x}_{2}\right)\\ \end{array}\right). \label{Eq00}
\end{eqnarray}
Besides the unit matrice (I), general position vectors $x_{1}$ and $x_{2}$, space-dependent Dirac matrices $\sigma _{\mu }^{\left( 1,2\right) }\left(x_{1},x_{2}\right)$, Kronocker productions are represented by $\otimes$ symbols in Eq. (\ref{Eq1}) where $e_{1}, e_{2}, m_{1}, m_{2}$ are charges and masses of the first and second particles, respectively.

Since we interest in only relative motion between the particles, to separate the center of mass and relative variables, the well known expressions are written in the following way;
\begin{eqnarray}
\mathbf{X}_{1}=\frac{m_{2}}{M}\left( \mathbf{r+}\frac{M}{m_{2}}\mathbf{R}
\right),\nonumber\\
\mathbf{X}_{2}=\frac{m_{1}}{M}\left(- \mathbf{r+}\frac{M}{m_{1}}\mathbf{R}
\right),\label{alg}
\end{eqnarray}
and their differential forms \cite{abdullah} are used.

\section{Coupled differential equation system}

For a Coulomb-type interaction, the components of electromagnetic vector potential, $A_{\mu}$, are written as;
\begin{eqnarray*}
\mathbf{A}_{0}=V\left\vert \mathbf{x}_{1}-\mathbf{x}_{2}\right\vert,\quad \mathbf{A}_{1}=0,
\end{eqnarray*}
and the space dependent matrices satisfying Dirac algebra can be chosen in terms of constant pauli spin matrices \cite{K} in the following way;
\begin{eqnarray}
\sigma _{0}=\sigma ^{z},\quad \sigma _{1}=i\sigma^{x}.\label{mat}
\end{eqnarray}

Thanks to separation of variables method, the components of the bi-spinor function written in Eq. (\ref{Eq00}) can be separated as,
\begin{eqnarray}
\xi _{y}\left( r,R,R_{0}\right):=\zeta _{y}\left(r\right)\Phi_{y}\left(R\right)e^{-iwR_{0}},\nonumber\\
\Phi_{y}\left( R\right):=e^{i\mathbf{k.R}},\ \ (y=1,2,3,4.),\label{SepS}
\end{eqnarray}
where $R_{0}$ and $R$ represent the proper time and spatial coordinate of the center of mass, respectively.

By using Eq. (\ref{SepS}), Eq. (\ref{mat}) and Eq. (\ref{alg}) in Eq. (\ref{Eq1}) an equation set is written as;
\begin{eqnarray*}
\lambda \left( r\right)\zeta _{a}\left( r\right)-B\zeta _{b}\left(
r\right)+ \left( 2\partial _{r}+\frac{\eta k}{M}\right)\zeta _{d}\left(
r\right) =0, \label{equation4}
\end{eqnarray*}
\begin{eqnarray*}
\lambda \left( r\right) \zeta _{b}\left( r\right) -B\zeta _{a}\left(
r\right) -k\zeta _{c}\left( r\right) =0, \label{equation5}
\end{eqnarray*}
\begin{eqnarray*}
\lambda \left( r\right) \zeta _{c}\left( r\right) -\triangle B \zeta
_{d}\left( r\right) +k\zeta _{b}\left( r\right) =0, \label{equation6}
\end{eqnarray*}
\begin{eqnarray}
\lambda \left( r\right) \zeta _{d}\left( r\right) -\triangle B \zeta
_{c}\left( r\right) - \left( 2\partial _{r}-\frac{\eta k}{M}\right) \zeta
_{a}\left( r\right) =0,  \label{equation7}
\end{eqnarray}
where we use the following abbreviations:
\begin{eqnarray}
\zeta _{a}\left( r\right)=\zeta _{1}\left( r\right) +\zeta _{4}\left(
r\right) , \ \zeta _{b}\left( r\right) =\zeta _{1}\left( r\right) -\zeta
_{4}\left( r\right) ,\nonumber\\
\zeta _{c}\left( r\right)=\zeta _{2}\left( r\right) +\zeta _{3}\left(
r\right) , \ \zeta _{d}\left( r\right) =\zeta _{2}\left( r\right) -\zeta
_{3}\left( r\right) ,\nonumber\\
\lambda \left( r\right)=\left( \frac{w}{c}-V\left( r\right) \right), V\left( r\right)=\frac{\alpha}{r}, \triangle B =\frac{\eta c}{\hbar},\nonumber\\
B=\frac{Mc}{\hbar}, \ \eta =m_{1}-m_{2},\ \ M=m_{1}+m_{2}.
\end{eqnarray}

\section{The spectra of a para-positronium in S-state}

Under the rest center of mass condition ($k=0$) and by adopting the equation set in Eq. (\ref{equation7}) to an electron-positron pair, $m_{1}=m_{2}$, a second order wave equation can be obtained as;
\begin{eqnarray}
\partial _{r}^{2}\zeta _{a}\left( r\right) -\left( \frac{\partial
_{r}\lambda \left( r\right) }{\lambda \left( r\right) }\right) \partial
_{r}\zeta _{a}\left( r\right) +\left( \frac{\lambda \left( r\right)
^{2}-B^{2}}{4}\right) \zeta _{a}\left( r\right) =0, \label{second}
\end{eqnarray}
and the following definition can be used to clarify solution functions of Eq. (\ref{second}),
\begin{eqnarray}
\zeta _{a}\left( r\right) =e^{\left(-\frac{r}{2 c}\sqrt{B^{2}c^{2}-w^{2}}
\right) }r^{\frac{i\alpha}{2}}\xi \left( r\right).\label{s}
\end{eqnarray}
Then, by changing dimensionless independent variable, as $z=-\frac{w r}{c\alpha}$, the solution functions of Eq. (\ref{second}), are found in terms of Heun functions,
\begin{eqnarray}
\xi \left( z\right) =A_{1}H_{C}\left( \theta ,\varepsilon ,\upsilon
,\varsigma ,\varrho ,z\right) +A_{2}H_{C}\left( \theta ,-\varepsilon
,\upsilon ,\varsigma ,\varrho ,z\right) z^{-\varepsilon },\label{s1}
\end{eqnarray}
where $\theta, \varepsilon, \upsilon, \varsigma$ and $\varrho$ are as follows;
\begin{eqnarray}
\theta &=&\frac{\alpha}{w}\sqrt{B^{2}c^{2}-w^{2}},\quad\varepsilon=-i\alpha,\nonumber\\
\upsilon&=&-2,\quad\varsigma =-\frac{\alpha^{2}}{2},\quad\varrho=1+\frac{\alpha^{2}}{2}.\label{s2}
\end{eqnarray}
With the help of following expression, which is the polynomial condition of the C-type Heun functions,
\begin{eqnarray}
\varsigma +\left( n+1+\frac{\left( \varepsilon +\upsilon \right) }{2}
\right) \theta =0,\label{s3}
\end{eqnarray}
the frequency spectra can be obtained as in the following;
\begin{eqnarray}
w_{n}=\frac{2m_{e}c^{2}}{\hbar}\sqrt{\frac{n^{4}+\frac{3\alpha^{2}n^{2}}{
4}-i\frac{\alpha^{3}n}{4}}{n^{4}+\alpha^{2}n^{2}}}.\label{fre}
\end{eqnarray}
The spectra including an imaginary part depends on the principal quantum number, ($n$), and well-known electromagnetic coupling constant ($\alpha$).
By using the power expansion method,
\begin{eqnarray}
w_{n}\approx \frac{2m_{e}c^{2}}{\hbar}\left\{ 1-\frac{\alpha ^{2}}{8n^{2}}-i\frac{\alpha ^{3}}{8n^{3}}\right\}, \label{spectrum}
\end{eqnarray}
the annihilation energy can be found as;
\begin{eqnarray}
\emph{E}_{ann}\approx \left(2m_{e}c^{2}-6,803\right) eV,\label{ann}
\end{eqnarray}
in which the $-6.803\quad eV$ is binding energy of the system in ground state ($-\frac{2m_{e}c^2\alpha^2}{8}\approx-6,803\quad eV$, n=1) and $2m_{e}c^{2}$ is the total rest mass energy of the system in vacuum.

Additionally, proper decay time of the system can be found, simultaneously, by using the expressions in Eq. (\ref{SepS}), which gives $\tau_{n}\approx\frac{1}{\left\vert imw_{n}\right\vert}$, as follows;
\begin{eqnarray}
\tau _{n}\approx \frac{4n^{3}\hbar}{m_{e}c^{2}\alpha ^{3}},\quad \alpha=\frac{e^{2}}{4\pi\epsilon_{0}\hbar c}.\label{lifetime}
\end{eqnarray}
For ground state, the proper decay time of system is found as; $\tau_{1}=0.0132\times10^{-12}\quad s$ in vacuum. This value is shorter than the observed lifetime values, even shorter than that of vacuum, since the obtained value corresponds only the proper lifetime of the system in ground state. However, the existence of any matter background, such as tissue or tumor, changes the binding energy, lifetime \cite{M,N} and total annihilation energy of the system due to the electronic properties and electron density of the sample matter. The electronic properties of the matter background can be represented by an effective dielectric constant, as $\epsilon_{m}$, and this term contains important physical properties of the environment of the annihilation position in the medium such as polarizability, cohesive energy density, temperature, permittivity or susceptibility \cite{Tao}.

\section{Conclusion}

The developed model in this paper gives exact annihilation energy, binding energy and proper decay time value, in S-state, of the p-Ps system, in vacuum. Since the obtained spectra showing fundamental properties of the p-Ps system is first spectrum in the literature, the yields can shed light to the related studies, such as positron annihilation spectroscopy for any sample material (or medium), medical monitoring of the living biological systems and gamma-ray laser studies based on the p-Ps annihilation process. The obtained proper decay time expression (Eq. (\ref{lifetime})), for vacuum case, gives the exact proper decay times for the system only in any S-state, therefore the obtained expressions can be based on to clarify the related results in the literature in event-by-event basis by taking into account well-known $\tau_{lab}$ correction \cite{S}. In any materials, after the system is formed, the external effects such as screening energy, substrate effect, an electromagnetic field or thermal fluctuations can change the whole spectrum of the Ps system. Since the screening energy or substrate effects in the medium impact on the electromagnetic coupling between the electron and positron pair as $\alpha\rightarrow\alpha_{m}, (\alpha_{m}<\alpha$), the proper decay time (Eq. (\ref{lifetime})), binding energy and total annihilation energy (Eq. (\ref{spectrum})) of the system are changed by the matter medium. These can be seen that the effects decrease the binding energy because the binding energy of the system is proportional to $\alpha_{m}^{2}$ term, but they increase the obtained proper decay time values as proportionally with the $\alpha_{m}^{-3}$ term. Hence, by sensitively measuring to the total annihilation energy transmitting by the two resulting photons via Doopler broadening spectroscopy \cite{dop} and the positron annihilation life time spectroscopy for any sample matter, the crucial physical properties, such as morphologic alteration and electronic environment of the annihilation position, in the sample can be understand. In this way, the obtained yields can contribute to studies about positron annihilation life time spectroscopy for any materials or living biological complexes and medical monitoring of  human tissues or cell membranes via positron emission tomography or multi-purpose Jagiellonian positron emission tomography devices. Thanks to the important developments in such areas, in near future, information about the cells, tissues or tumors may be quitely detected via sensitive dedectors used in the medical monitoring devices \cite{O}. Finally, because of the total annihilation energy and decay time of the p-Ps system depend explicitly on the medium properties, our findings can be based on for gamma-ray laser studies \cite{laser}, since the formation time, annihilation time and total annihilation energy transmitting by the two annihilation gamma photons propagating as forward photon and backward photon in the time may be controlled by arranging the matter substrate sensitively.

\end{document}